\documentclass[10pt,aps,prd,twocolumn,showpacs,superscriptaddress,nofootinbib,nobibnotes,longbibliography,floatfix]{revtex4-1}

\usepackage[utf8]{inputenc}
\usepackage[T1]{fontenc}
\usepackage{enumitem}
\usepackage{bm}
\usepackage{mathtools,amsmath,amssymb,amsfonts,mathrsfs,eucal,graphicx,tensor,csquotes,accents, commath,chngcntr,siunitx}
\usepackage[dvipsnames]{xcolor}
\usepackage[unicode]{hyperref}
\hypersetup{colorlinks=true, citecolor=MidnightBlue,
            linkcolor=MidnightBlue, urlcolor=MidnightBlue, linktocpage=true}
\usepackage[normalem]{ulem}
\usepackage{amsmath}
\usepackage{tensor}

\DeclareMathAlphabet{\mathpzc}{OT1}{pzc}{m}{it}
\definecolor{darkgreen}{rgb}{0.0, 0.6, 0.0}

\begin{document}

\author{Lodovico Capuano}
\affiliation{SISSA, Via Bonomea 265, 34136 Trieste, Italy and INFN Sezione di Trieste}
\affiliation{IFPU - Institute for Fundamental Physics of the Universe, Via Beirut 2, 34014 Trieste, Italy}

\author{Luca Santoni}
\affiliation{Universit\'e Paris Cit\'e, CNRS, Astroparticule et Cosmologie, 10 Rue Alice Domon et L\'eonie Duquet, F-75013 Paris, France}

\author{Enrico Barausse}
\affiliation{SISSA, Via Bonomea 265, 34136 Trieste, Italy and INFN Sezione di Trieste}
\affiliation{IFPU - Institute for Fundamental Physics of the Universe, Via Beirut 2, 34014 Trieste, Italy}

\title{Black hole hairs  in scalar-tensor gravity and the lack thereof}

\begin{abstract}
Scalar-tensor theories are a natural alternative to general relativity, as they may provide an effective dark energy phenomenology on cosmological scales while passing local tests, but their black hole solutions are still poorly understood. Here, we generalize existing no-hair theorems for spherical black holes and specific theories in the scalar-tensor class. We show that  shift symmetry prevents the appearance of scalar hairs in rotating (asymptotically flat, stationary and axisymmetric) black holes for all theories in the Horndeski/beyond Horndeski/DHOST classes, but for those with a coupling between the scalar and the Gauss-Bonnet invariant. Our proof also applies to higher dimensions. We also compute the values of the scalar hair charges if shift symmetry and asymptotic flatness are violated by a time growth of the scalar field at infinity, under suitable regularity conditions at the event horizon.
\end{abstract}

\maketitle

\section{Introduction}  
Scalar-tensor theories of gravity are the simplest and oldest extension of general relativity (GR). In their basic form, they date back to the pioneering work by Fierz~\cite{Fierz:1956zz}, Jordan~\cite{Jordan:1959eg}, Brans and Dicke~\cite{Brans:1961sx}, who suggested supplementing the tensor gravitons of GR with a scalar degree of freedom conformally coupled to matter.
This Fierz-Jordan-Brans-Dicke (FJBD) theory, and generalizations of it
in which the conformal coupling to matter is expanded to nonlinear orders~\cite{Damour:1992we,Damour:1993hw},
have been for decades the paradigmatic extensions of GR when performing experimental tests in the Solar System~\cite{Will:1993hxu} and binary pulsars~\cite{Damour:1991rd}. This has resulted in very tight bounds on these theories~\cite{Will:2014kxa}. 

The observational evidence for a dark sector in cosmology~\cite{Astier:2012ba,SupernovaCosmologyProject:1998vns,Planck:2015bue, Planck:2018vyg}
and the direct detection of gravitational waves by LIGO and Virgo~\cite{LIGOScientific:2016aoc} 
have spurred a resurgence of interest in scalar-tensor theories. It was realized that FJBD-like theories are not the most general ghost-free theories allowing for a scalar degree of freedom in addition to the tensor ones.
Indeed, FJBD-like theories are just a special case of more general effective field theories (EFTs), where all possible scalar-tensor operators are organized in a derivative expansion. Higher derivative operators typically provide subleading corrections on a given solution at low energies, however there are cases where they can be as important as the ones with fewer derivatives, within the domain of validity of the low energy expansion. This property can be made robust by the presence of exact or approximate symmetries,  which determine different sets of power-counting rules for the coupling constants in the effective Lagrangian, even in the absence of an explicit UV completion~\cite{Nicolis:2008in,Pirtskhalava:2015nla,Santoni:2018rrx}. The simplest realization of these effective theories belongs to the Horndeski class~\cite{Horndeski:1974wa}, described by the following action\footnote{In Eq.~\eqref{horndeski} we are not explicit about the energy scales associated with the derivative operators, which are absorbed in the definitions of $K$ and $G_i$;  we will restore them later on when needed.}
\begin{eqnarray}
\label{horndeski}
&&S=\int {\rm d}^4x \frac{\sqrt{-g} M_{\rm Pl}^2}{2}\Big\{ K(\psi,X) -G_3(\psi,X)\Box\psi \\&&+ G_{4}(\psi,X)R
+G_{4X}(\psi,X)\left[
\left(\Box\psi\right)^2-\left(\nabla_\mu\nabla_\nu\psi\right)^2
\right]\nonumber
\\
&&+G_5(\psi,X) G_{\mu\nu}\nabla^\mu\nabla^\nu\psi
-\frac{G_{5X}(\psi,X)}{6}\Bigl[
\left(\Box\psi\right)^3
\nonumber\\&&
-3\left(\Box\psi\right)\left(\nabla_\mu\nabla_\nu\psi\right)^2
+2\left(\nabla_\mu\nabla_\nu\psi\right)^3
\Bigr]\Big\}+S_m[g_{\mu\nu},\Psi_m]\nonumber
\end{eqnarray}
where $g$, $\nabla$, $R$ and $G_{\mu\nu}$ and $M_{\rm Pl}$ are the metric determinant,  Levi-Civita connection, Ricci scalar, Einstein tensor and reduced Planck mass;
$K$, $G_3$, $G_4$, and $G_5$ are arbitrary functions of $X\equiv-\nabla_\mu\psi \nabla^\mu\psi/2$ 
and the scalar field $\psi$;
$G_{iX}\equiv \partial G_i/\partial X$, 
$\Box \equiv \nabla^\mu\nabla_\mu$, $\left(\nabla_\mu\nabla_\nu\psi\right)^2  \equiv \nabla_\mu\nabla^\nu \psi\nabla_\nu\nabla^\mu\psi$ and
$\left(\nabla_\mu\nabla_\nu\psi\right)^3 \equiv \nabla_\mu\nabla^\rho \psi\nabla_\rho\nabla^\nu\psi\nabla_\nu\nabla^\mu\psi$; 
and $\Psi_m$  are the matter fields. 
The class of theories given by Eq.~\eqref{horndeski} can be further generalized  to the beyond Horndeski~\cite{Gleyzes:2014dya,Zumalacarregui:2013pma} and degenerate-higher-order-scalar-tensor (DHOST) theories~\cite{Langlois:2015cwa, Langlois:2015skt, Crisostomi:2016tcp, BenAchour:2016fzp, BenAchour:2016cay, Crisostomi:2016czh} (see \cite{Langlois:2018dxi} for a review).
The latter is defined as the most general class of scalar-tensor theories with no propagating ghost degrees of freedom, although only a subset of those can be considered as ``robust'' EFTs~\cite{Pirtskhalava:2015nla,Santoni:2018rrx,Noller:2019chl}. 

The coincident detection of gravitational waves and gamma rays from the neutron star merger GW170817~\cite{Monitor:2017mdv}, as well as the requirement
that gravitational waves do not decay into dark energy  \cite{Creminelli:2018xsv,Creminelli:2019nok}
and that the scalar mode be nonlinearly stable \cite{Creminelli:2019kjy}, have already placed very strong constraints on DHOST, under the assumption that the theory provides a dark energy like phenomenology on cosmological scales. With this assumption, the only theories still viable  are described  by the action~\cite{Lara:2022gof}
\begin{eqnarray}
&& S=\!\!\int {\rm d}^4 x \frac{\sqrt{-g} M_{\rm Pl}^2}{2}  \Bigg[\Phi \,R +K(\psi,X) \label{Jframe_action} \\
&& \qquad +\frac{3\, \Phi_X^2}{2\,\Phi} \nabla^{\mu}\psi \nabla_{\mu \rho}\psi \nabla^{\rho \nu}\psi \nabla_{\nu}\psi\Bigg] + S_m[g_{\mu\nu},\Psi_m]\,, \nonumber
\end{eqnarray}
where $\Phi$ and $K$ are functions of $\psi$ and $X$ (with $\Phi_X \equiv \partial \Phi/\partial X$).
With a  conformal transformation from the ``Jordan frame'' to the ``Einstein frame'', i.e.~$g_{\mu\nu} \to \Phi^{-1}\, g_{\mu\nu}$, the action can be rewritten (redefining the function $K$) as the ``$K$-essence'' action

 \begin{equation}
\!S=\!\!\int\!{\rm d}^4 x\sqrt{-g}\Bigg[ \frac{M_{\rm Pl}^2}{2}R + K(\psi,X)
\Bigg] \! +S_m\!\!\left[\frac{g_{\mu\nu}}{\Phi(\psi,X)},\Psi_m\right].
\label{eq: kessence action}
\end{equation}

The conformal coupling  to matter $\Phi(\psi,X)$
can potentially be tested with observations of neutron stars (in isolation~\cite{Damour:1993hw} or in binaries~\cite{Barausse:2012da,Palenzuela:2013hsa}). The kinetic function $K(\psi,X)$ also has important consequences for the dynamics of matter systems. For instance, specific kinetic functions can give rise to self-accelerated solutions in cosmology~\cite{Armendariz-Picon:2000nqq}, or to nonlinear screening mechanisms that ``hide'' the deviations from GR on local scales (at least in quasistatic situations~\cite{Babichev:2009ee,terHaar:2020xxb,Bezares:2021yek,Bezares:2021dma}). However, the cleanest probes of the kinetic term $K(\psi,X)$ are provided by vacuum systems (e.g.~black holes), since for those the effect of the conformal coupling function $\Phi(\psi,X)$ vanishes. This is particularly interesting in light of the several black binary systems detected by LIGO and Virgo~\cite{LIGOScientific:2021djp}.

The deviations of the gravitational-wave signal (and more generally of the geometry) of black holes in scalar-tensor theories from their GR counterparts is parametrized in terms of scalar hairs (also referred to as sensitivities~\cite{osti_4238685,Will:1989sk} or scalar charges~\cite{Damour:1992we,Damour:1993hw}). These parameters model the effective coupling between black holes and the scalar graviton in these theories, and are therefore absent in GR. In fact, these charges can also be thought of as quantifying violations of the no-hair theorem~\cite{Kerr:1963ud, Newman:1965my, Carter:1968rr, PhysRevLett.34.905} and of  the strong equivalence principle~\cite{Will:1993hxu}, which are satisfied in GR but not necessarily in more general gravitational theories.

The appearance of scalar hairs, however, is not inevitable. In FJBD-like theories,
with or without a scalar field mass, no-hair theorems exist and dictate that black hole solutions must match the GR ones
 if one assumes asymptotically flat boundary conditions~\cite{cmp/1103857885}.\footnote{Note that introducing a matter content, one can derive also no-hair theorems for stars \cite{Star_NoHair_Horndeski,Challenging_Scalar_Charge}.} This theorem also applies to black holes in $K$-essence, under the same asymptotically flat boundary conditions and provided that the kinetic function satisfies suitable ``stability'' conditions~\cite{NoHair_KEssence}. 

Black hole hairs, however, generically appear, even in $K$-essence and FJBD-like theories, if the scalar field grows with time far from the black hole~\cite{Jacobson_1999, Charmousis:2019vnf, Stealth_Hairs, Van_Aelst_2020}, as would be expected if one were to match to a cosmological solution on large scales. Moreover, if one does not require  the scalar field to provide an effective dark energy phenomenology, the aforementioned bounds on the DHOST class (coming from gravitational-wave propagation, the decay of gravitational waves into the scalar mode, and the nonlinear stability of the latter) are no more applicable. No-hair theorems exist for subsets of the DHOST class in spherical symmetry~\cite{Hui:2012qt}, but they rely on shift symmetry and on the assumption that the free functions appearing in the DHOST action are analytic.
Therefore, not only  are scalar charges  generically expected for DHOST theories that break shift symmetry in vacuum, but these theorems also do not apply to 
actions including interactions between the scalar field and the Gauss-Bonnet invariant~\cite{sGB_4D, GaussBonnetGravity_Review} (which corresponds to a nonanalytic $G_5\propto \ln |X|$~\cite{General_Shift_Symmetric,General_Shift_Symmetric_2}). These couplings
 are known to produce black hole charges that can be even nonperturbatively large (black hole scalarization) in both the spherical~\cite{Dilaton_BH, sGB_Gravity_Solutions} and rotating~\cite{Spin_Induced_Scalarization_Herdeiro,Spin_Induced_SpontaneousScalarization} case.

 In this paper, we generalize  existing no-hair theorems in the context of scalar-tensor gravity and review the situations where these theorems can be violated. 
 In more detail, in Sec. \ref{SSST_Stationary}, we provide the proof of a no-hair theorem for stationary asymptotically flat black holes, holding for any shift-symmetric scalar-tensor theory (including Horndeski and DHOST). In Sec. \ref{higher_d}, we show that this result can be generalized to an arbitrary number $d$ of spacetime dimensions, under suitable assumptions on the topology of the horizon. In Sec. \ref{no_hair_violation}, we  
 provide examples which violate the no-hair theorems, explaining which assumptions of our proof are violated. We also compute the scalar charge associated with a linear time dependence of the scalar field, under suitable regularity conditions at the event horizon. Finally, in Sec. \ref{Quasi_Kessence}, we further comment on possible extensions of the existing theorems dropping the shift-symmetry assumption. 
 
\textit{Notation and conventions:}
We will work in natural units $c = \hbar = 1$ and in the  mostly plus signature of the metric, $(-,+,+,\dots)$. In the main text, we will use greek indices $\mu,\nu,\alpha,\dots$ for the spacetime coordinates on the $d$ dimensional geometry,  lowercase latin indices $i,j,k,\dots$ (from the middle of the alphabet) for the $d-1$ spatial coordinates only, and lowercase latin indices $a,b,c,\dots$ (from the beginning of the alphabet) to label the angular coordinates associated with the rotational Killing vectors of the metric (see e.g.~Sec.~\ref{higher_d}). (A slightly different convention is adopted though in Appendix~\ref{proof_Jt_0}.)

%%%%%%%%%%%%%%%%%%%%%%%%%%%%%%%%%%%%%%%%%%%%%%%%%%%%%%%%
\section{No-hair theorem for rotating black holes}
\label{SSST_Stationary}
In this section, we  
prove a no-hair theorem for rotating black holes
in
four spacetime dimensions, 
which we generalize to
arbitrary spacetime dimensions  in Sec.~\ref{higher_d}. In particular, we  show that asymptotically flat, axisymmetric and stationary black holes in shift-symmetric scalar-tensor theories cannot develop a term
$\propto 1/r$ in the scalar profile at large distances. This is equivalent to saying that black holes cannot have a scalar charge~\cite{Damour:1992we,Damour:1993hw}.
In the language of the point-particle effective theory, this corresponds to a vanishing hair coupling $g$ in the tadpole operator $g \psi$ localized on the worldline of the point particle; see e.g.~\cite{Goldberger:2004jt,Porto:2016pyg,Hui:2020xxx}.

The fundamental assumptions that we make are the following:
\begin{enumerate}[label=(\roman*)]
\item the metric is circular; i.e.~it has two commuting Killing vectors associated respectively to the invariance under shifts in the time coordinate $t$ (stationarity) and in the azimuthal angle $\phi$ (axisymmetry), and it is invariant under the reflection isometry $\{t\rightarrow-t,\,\phi\rightarrow-\phi\}$;
\item the spacetime is asymptotically flat, reducing at large radii to the Minkowski metric $\eta_{\mu\nu}$, plus subleading corrections $h_{\mu\nu}\sim \mathcal{O}(1/r)$;
\item The action and the field equations for the scalar field $\psi$ are invariant under  shifts  $\psi\rightarrow\psi+c$, with $c$ a constant;
\item if a nontrivial solution for the scalar field exists, it has the same symmetries as the spacetime, i.e., it does not depend on the coordinates associated with the Killing vectors of the metric;
\item the squared norm $J^\mu J_\mu$ of the conserved Noether current $J_\mu$ associated with the shift symmetry $\psi\rightarrow\psi+c$ is regular at the horizon;
\item the current $J_\mu$ reduces asymptotically to that of a free massless scalar field, i.e.~$J_\mu=-\partial_\mu\psi$ at large distances.
\end{enumerate}
Note that the latter condition is expected to hold in any standard effective theory for a scalar coupled to gravity, where derivative interactions computed on the background are more suppressed by powers of $1/r$ at large distances.\footnote{In particular, it holds for the Horndeski Lagrangian (\ref{horndeski}) and more in general for the DHOST class \cite{Langlois:2015cwa, Langlois:2015skt, Crisostomi:2016tcp, BenAchour:2016fzp, BenAchour:2016cay, Crisostomi:2016czh}.} 
To be concrete, we will sometimes refer below to the explicit Horndeski action in Eq.~(\ref{horndeski}), with shift current  $J_{\mu}$ given by 
\begin{equation}
\begin{split}
    J_{\mu}= &-\nabla_{\mu}\psi\,\Big\{K_X-G_{3X}\square\psi\\
    &+G_{4X}R+G_{4XX}\big[(\square \psi)^2-(\nabla_{\alpha}\nabla_{\beta}\psi)^2\big]\\
    &+G_{5X}G^{\alpha\beta}\nabla_{\alpha}\nabla_{\beta}\psi-G_{5XX}\big[(\square \psi)^3\\
    &-3\square \psi\,(\nabla_{\alpha}\nabla_{\beta}\psi)^2+2(\nabla_{\alpha}\nabla_{\beta}\psi)^3\big]\Big\}\\
    &-\partial^{\nu}X\Big\{-g_{\mu\nu}\,G_{3X}\\
    &+2G_{4XX}(\square \psi \,g_{\mu\nu}-\nabla_{\mu}\nabla_{\nu}\psi)\\
    &+G_{5X}G_{\mu\nu}-\frac{1}{2}G_{5XX}\big[g_{\mu\nu}(\square \psi)^2\\
    &-g_{\mu\nu}( \nabla_{\alpha}\nabla_{\beta}\psi)^2\\
    &-2\square \psi\, \nabla_{\mu}\nabla_{\nu}\psi+2 \nabla_{\mu}\nabla_{\sigma}\psi\nabla^{\sigma}\nabla_{\nu}\psi\big]\Big\}\\
    &+2G_{4X}R_{\mu\sigma}\nabla^{\sigma}\psi +G_{5X}(-\square \psi \,R_{\mu\sigma}\nabla^{\sigma}\psi\\
    &+\tensor{R}{^\alpha_\nu^\beta_\mu}\nabla_{\alpha}\nabla_{\beta}\psi\nabla^{\nu}\psi+\tensor{R}{_\alpha^\beta}\nabla^{\alpha}\psi\nabla_{\mu}\nabla_{\beta}\psi)\,.
\label{Horndeski_current}
\end{split}
\end{equation}
However, our result is more general, as it relies only on  assumptions (i)-(vi), and does not depend on the explicit form of the (shift-symmetric) scalar action. We will discuss in Sec.~\ref{Time_Dependence} the possibility of relaxing some of the assumptions above.

\begin{figure}[h]
          \centering
          \includegraphics[width =0.27\textwidth]{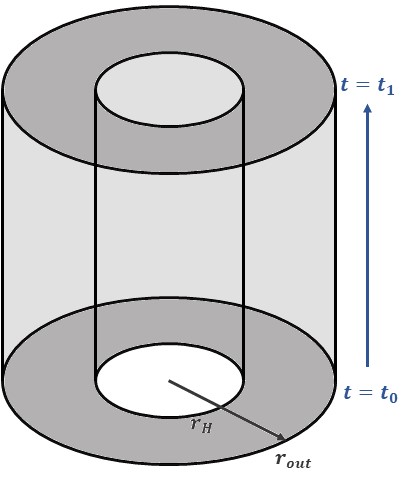}
          \caption{Schematic representation of the boundary $\partial\mathcal{M}$ appearing in Eq.~\eqref{Stokes}.}
          \label{fig:cylinder}
\end{figure}

Let us start by noting that, thanks to the shift symmetry, the scalar's equations of motion can be expressed, in  absence of matter, in the form of a (covariant) conservation law,
\begin{equation}
    \label{current_conservation}
    \nabla^{\mu}J_{\mu} = 0 \,,
\end{equation}
where $\nabla_{\mu}$ is the covariant derivative. 
Integrating Eq.~\eqref{current_conservation} over the spacetime outside the horizon and using Stokes' theorem, one gets
\begin{equation}    
    \oint_{\partial\mathcal{M}} {\rm d} \Sigma_{\mu}\,g^{\mu\nu}\,J_{\nu}=0\,,
\label{Stokes}
\end{equation}
where $\partial\mathcal{M}$ is the  three-dimensional boundary  of the black hole exterior region and ${\rm d} \Sigma_{\mu}$ is the element of the hypersurface $\partial\mathcal{M}$.
Introducing a radial coordinate $r$ constant on the horizon ($r=r_{\rm H}$) - which can be done without loss of generality if we assume that the horizon has the topology of a sphere \cite{1973blho.conf..241B} - and a time coordinate $t$, the boundary  $\partial\mathcal{M}$  includes four contributions: two with fixed radius ($r=r_{\rm H}$ or
$r=r_{\rm out}\to\infty$) and variable $t\in [t_0,t_1]$
(with $t_0$ and $t_1$ two constants), and two with $t=t_0$
or $t=t_1$ and $r\in [r_{\rm H},r_{\rm out}]$.
See Fig.~\ref{fig:cylinder} for a sketch of the domain of integration.  

Let us focus first on the horizon contribution to the boundary integral in Eq.~\eqref{Stokes}.
The Cotton-Darboux theorem~\cite{darboux1910lecons,chandrasekhar1992mathematical} ensures that  it is always possible to recast the metric of a  three-manifold into diagonal form via a local coordinate transformation. In particular, in $d=4$ dimensions, under the assumptions of stationarity and axisymmetry, we can always choose coordinates such that the line element for the exterior geometry takes the Weyl-Papapetrou  form \cite{PhysRevLett.26.331,Bekenstein_II}:
\begin{equation}
       ds^2 = -P\,{\rm d}t^2+ 2 Q\,{\rm d}t\,{\rm d}\phi+H\,{\rm d}\phi^2
       +W\,\big({\rm d}\rho^2+{\rm d}z^2\big),\label{metric_ansatz}
\end{equation}
 where the only off diagonal component of the metric is $g_{t\phi}=Q$.
As a result, ${\rm d } \Sigma_t= {\rm d } \Sigma_{\phi}={\rm d } \Sigma^t= {\rm d } \Sigma^{\phi}=0$ on the horizon hypersurface $r=r_{\rm H}$.
In addition, using the assumptions (iii)-(iv), one can show in general that $J_t= J_{\phi}=0$, and thus also $J^t= J^{\phi}=0$, everywhere, and in particular at the horizon. This can be checked explicitly for the Horndeski current in Eq.~(\ref{Horndeski_current}), while a general proof holding for any theory that satisfies the above requirements can be found in Appendix \ref{proof_Jt_0}. 
Thus, we can write
\begin{equation}  
 {\rm d} \Sigma_{\mu}\,g^{\mu\nu}\,J_{\nu} = {\rm d}\Sigma_i g^{ij} J_j
\label{horterm}
\end{equation}
 as an inner product in a three-dimensional space. 

Using Eq.~\eqref{metric_ansatz}, one obtains, more explicitly,
\begin{equation}
\begin{split}
     ({\rm d}\Sigma_{i}\,g^{ij}\,J_{j})^2 & = W^{-2}\big({\rm d} \Sigma_\rho J_\rho +{\rm d} \Sigma_z J_z \big)^2 \\
     &\leq W^{-2}\big({\rm d} \Sigma_\rho^2+{\rm d} \Sigma_z^2\big)\,\big(J_\rho^2+J_z^2\big) = \\&= (J_{i}\,g^{ij}\,J_{j})\,({\rm d}\Sigma_{k}\,g^{kl}\,{\rm d}\Sigma_{l})\,,
\end{split}
\label{Cauchy_Schwarz_4d}
\end{equation}
where we have used the Cauchy-Schwarz inequality. 
From the regularity of $J_{i}\,g^{ij}\,J_{j}$, which follows from the assumption (v) and the fact that $J_t=J_\phi=0$, and from the vanishing of ${\rm d}\Sigma_{k}\,g^{kl}\,{\rm d}\Sigma_{l}$ at $r=r_{\rm H}$ by definition of null hypersurface, it follows that the right-hand side of Eq.~(\ref{Cauchy_Schwarz_4d}) vanishes at the horizon. 
As a consequence, the left-hand side of Eq.~\eqref{Cauchy_Schwarz_4d} must vanish as well. This proves that in $d=4$  there is no flux at the horizon contributing to the integral in Eq.~(\ref{Stokes}).\footnote{The vanishing of this contribution can also be proven, in $d=4$ dimensions and for Horndeski theories,
by assuming that the
surface gravity of the horizon is constant and the scalar field is regular~\cite{Saravani_2019,Benkel_2018}.}

Let us now focus on the constant-time boundaries of the region in Fig.~\ref{fig:cylinder}.
On the hypersurfaces $t=t_0$, or $t=t_1$, one has ${\rm d}\Sigma_{\mu}\propto \delta^t_{\mu}$ \cite{Bekenstein_II}. Therefore, since $J_t=J_\phi=0$, ${\rm d}\Sigma_\mu\,g^{\mu\nu}\,J_\nu=0$ at both $t=t_0$ and $t=t_1$.\footnote{It is actually not necessary that $J_t$ and $J_\phi$ vanish. It is enough that they are 
independent of $t$ and $\phi$, as required by stationarity and axisymmetry. In fact, ${\rm d}\Sigma_\mu$ points in opposite directions at  $t=t_0$ and $t=t_1$;  thus, for $J_t$ and $J_\phi$ independent of $t$ and $\phi$, the fluxes at the time boundaries are guaranteed to cancel each other.}

We are  therefore left  only with the boundary contribution  at large radii. Choosing the boundary to be at  $r=r_{\rm out}\rightarrow\infty$, the flux reads
\begin{equation}
\oint_{\partial\mathcal{M}} {\rm d} \Sigma_{\mu}\,g^{\mu\nu}\,J_{\nu}=    \int_{\{r\rightarrow\infty\}}{\rm d} \Sigma_r\,g^{rr}\,J_r\,.
\label{integral_infinit_sphere}
\end{equation} 

At large distances, we can express the scalar profile as $\psi=\sum_{\ell} \psi_\ell(r) P_\ell(\cos\theta)$,
where $P_\ell$ are the Legendre polynomials and $\psi_\ell$ (because of asymptotic flatness) go at leading order as\footnote{We are using here the assumption that the scalar equations of motion reduce, at large distances from the black hole, to those of a free scalar field, and we are disregarding the other independent solution with falloff $\sim r^{\ell}$.}
\begin{equation} 
   \psi_\ell \sim \frac{a_\ell}{r^{\ell+1}} \, ,
\label{MultipolarExp}
\end{equation} 
for constant $a_\ell$.
In particular, the coefficient $a_0$ of the monopole term with $\ell=0$ is related to the scalar charge, or "hair", $Q_S$ via $a_0 = Q_S GM$, where $M$ is the black hole mass \cite{Bezares:2021yek, Constraints_GB_Gravity, Palenzuela:2013hsa}.
Let us now use  assumption (vi) and write \cite{Damour:1991rd,Damour:1992we}
\begin{equation}
J_r \sim -\partial_r \psi\,
\label{asymptotic_current}
\end{equation}
at large distances from the black hole. Furthermore, from assumption (ii) we have $g^{rr}=1+\mathcal{O}(1/r)$.
Computing the integral (\ref{integral_infinit_sphere}) using spherical polar coordinates one gets

\begin{equation}
\begin{split}
    \int_{\{r\rightarrow\infty\}}{\rm d} \Sigma_r\,g^{rr}\,J_r 
    &=\lim_{r \to \infty} r^{2}\int {\rm d} t \, {\rm d}\Omega_{S^{2}}\,J_r  \\
    &=(t_1-t_0)\,4\pi\,G M Q_S =0\,,
\end{split}
\label{ScalarChargeVanishing}
\end{equation} 
which implies $Q_S=0$. 

Hence, we have  shown that under the  assumptions (i)-(vi) above, the scalar charge is always zero in four dimensions. This result is independent of the particular form of the (shit-symmetric) scalar action and can be generalized to arbitrary dimensions, as we will discuss in the next section. In this sense, it is a generalization of the no-hair theorem of Ref.~\cite{Hui:2012qt} - which applies to non-rotating black holes in shift-symmetric scalar-tensor theories - as well as of the theorem of Refs.~\cite{Bekenstein_I,Bekenstein_II,NoHair_KEssence} - which applies to rotating black holes,  but which holds only in theories with at most single derivatives acting on the scalar field. However, our result is weaker because it rules out only the monopole scalar hair, but does not exclude the possibility of nonvanishing subleading multipole terms in the scalar profile.

%%%%%%%%%%%%%%%%%%%%%%%%%%%%%%%%%%%%%%%%%%%%%%%%%%%%%%%%%
\section{Generalization to higher dimensions}
\label{higher_d}
We will now  generalize the results of the previous section to  arbitrary spacetime dimensions $d>4$.
The landscape of vacuum solutions in higher-dimensional GR is richer than in four dimensions. In $d>4$ there exist black objects with extended horizons,  such as black strings and black $p$-branes, as well as solutions presenting horizons with nontrivial topology, such as black rings. See, e.g., Ref.~\cite{Emparan:2008eg} for a review.
For simplicity, we will focus here on spacetimes with horizons that have the topology of a sphere, and restrict our analysis to the class of Myers-Perry black holes (with single or multiple spins) \cite{Myers_Perry,Emparan:2008eg}, leaving the generalization to different types of solutions  for future work.

The  fundamental assumptions will be the same as in the previous section, properly generalized. In particular, we will assume that (i) the metric has $n+1$ commuting Killing vectors given by $\partial_t$ and $\partial_{\phi_a}$, for $a=1,\dots, n$, with $n\leq d-2$, and is invariant under the symmetry transformation $\{t\rightarrow-t,\phi_1\rightarrow-\phi_1,\dots,\phi_n\rightarrow-\phi_n\}$.
%\footnote{We keep the number of axial Killing vectors general, even though one would be enough for the proof.}
In addition, we require (ii) asymptotic flatness, with  subleading corrections scaling now as $1/r^{d-3}$ for $h_{\mu\nu}$ and $\psi$. 
The assumptions (iii)-(vi) of Sec.~\ref{SSST_Stationary} are instead unchanged, with the only obvious remark that (iv) should be now understood to apply to all $n+1$ Killing vectors of the metric.

Note that Eqs.~\eqref{current_conservation} and \eqref{Stokes} (where now $\partial\mathcal{M}$ is the  generalization of the boundary hypersurface in Fig.~\ref{fig:cylinder} to higher dimensions) hold in any spacetime dimension. Therefore, the first part of the argument is identical up to Eq.~\eqref{Stokes}. Let us thus reconsider the different contributions to the flux in Eq.~\eqref{Stokes}. First, notice that the contribution from the  boundaries at fixed $t_0$ and $t_1$ vanishes for the same reason discussed above.  
Regarding the boundary at $r = r_H$, one can define the null directed surface element $\rm d \Sigma_\mu$ orthogonal to the horizon, where ${\rm d} \Sigma_t = {\rm d} \Sigma_{\phi_a}=0$ because of (i). Hence Eq.~\eqref{horterm} also remains true.
However, showing that $g^{ij}$ is positive-definite requires slightly adjusted considerations, because one cannot generically guarantee that $g^{ij}$
can be put in diagonal form.
We shall thus proceed as follows.
First, from the positivity of the spatial line element ${\rm d}l^2 = g_{ij}{\rm d } x^i {\rm d } x^j $, where latin indices here denote the $n-1$ spatial coordinates, it follows that  $g_{ij}$ with lower spatial indices is  positive-definite everywhere in the black hole exterior. 
Furthermore, one can choose the  coordinates in such a way that the only off diagonal terms in the metric tensor are those that mix time with the angles $\phi_a$ associated with the spin direction(s). In reference to this, see e.g.~the explicit form of the Myers--Perry line element in Appendix \ref{Kerr_d_dim}. Hence, the metric can be expressed as
\begin{equation}
    \left(g_{\mu\nu}\right) = \begin{pmatrix}
    g_{tt} & u_j\\
    u_i & \gamma_{ij}
    \end{pmatrix}\,,
\label{gdecomp}
\end{equation}
where the only nonvanishing components of the vector $u$ are those corresponding to the coordinates $\phi_i$, and where $\gamma$ is a positive-definite $(d-1)\times(d-1)$ matrix, defined as $\gamma_{ij}= g_{ij}$. Note that the spatial indices of the metric blocks are raised/lowered as 
\begin{equation}
 u^i = \gamma^{ij}u_j\,,\qquad\qquad
  \gamma_{ik}\gamma^{kj}=\delta_{i}^{j}  \,.
\end{equation}
Using the inversion rule for a block matrix, we get  the spatial part of the inverse metric as
\begin{equation}
    g^{ij}=\gamma^{ij}+\frac{\gamma^{ik}\,u_k\,u_l\,\gamma^{lj}}{g_{tt}-u_k\,\gamma^{kl}\,u_l}\,.
\label{Sherman_Morrison}
\end{equation}
From the symmetries of the black hole solution,
it follows that $\gamma^{-1}$ is block diagonal and does not mix the $\phi_a$ directions with the other spatial coordinates. Therefore, ${\rm d}\Sigma_i \gamma^{ik}\,u_k=J_i \, \gamma^{ik}\,u_k=0$. Furthermore, ${\rm d} \Sigma_t = {\rm d} \Sigma_{\phi_a}=J_t=J_{\phi_a}=0$ (see Appendix~\ref{proof_Jt_0}). Hence, in analogy with the $d=4$ case, we have

\begin{equation}
    {\rm d}\Sigma_{\alpha}\,g^{\alpha\beta}\,J_{\beta} = {\rm d}\Sigma_i\,\gamma^{ij}\,J_j \,.
\end{equation}
At this point, since $\gamma^{-1}$ is positive definite, we can safely apply the Cauchy--Schwarz inequality 
\begin{equation}
     ({\rm d}\Sigma_{i}\,\gamma^{ij}\,J_{j})^2 \leq (J_{i}\,\gamma^{ij}\,J_{j})\,({\rm d}\Sigma_{k}\,\gamma^{kl}\,{\rm d}\Sigma_{l})\,.
\label{Cauchy_Schwarz}
\end{equation}
From the regularity of  $J^\mu J_\mu$ at the horizon [assumption (v)] and $J_t=J_{\phi_a}=0$, the right-hand side of Eq.~\eqref{Cauchy_Schwarz} is zero, as ${\rm d}\Sigma$ is null and has therefore vanishing norm at the horizon.\footnote{
Actually, one needs to apply the argument
slightly away from the horizon, and a limit must be  taken.}
This shows that the contribution to the  integral \eqref{Stokes} from this boundary is zero, just like in the $d=4$ case. 

At large distances, we can decompose the scalar field as $\psi=\sum_{L} \tilde\psi_L(r) Y_L(\theta)$, where $Y_L$ are hyperspherical harmonics, $\theta$ is a shorthand for the angles on the $(d-2)$-dimensional hypersphere (see, e.g., Ref.~\cite{Hui:2020xxx}),
and $\psi_L$ scale at leading order as
\begin{equation} 
\tilde \psi_L \sim \frac{a_L}{r^{L+d-3}} \, ,
\label{MultipolarExp_d}
\end{equation}
for constant $a_L$.
The scalar charge can be defined via $a_0 = Q_S \,\mu$, where $\mu$ is the $d$-dimensional mass parameter defined in Appendix~\ref{Kerr_d_dim}, which reduces to $2GM$ in four dimensions. 

Computing the flux at spatial infinity yields

\begin{equation}
\begin{split}
    \int_{\{r\rightarrow\infty\}}{\rm d}\Sigma_{r}\,g^{rr}\,J_r 
    &=\lim_{r \to \infty} r^{d-2}\int {\rm d} t \, {\rm d}\Omega_{S^{d-2}} \,g^{rr}\,J_r  \\
    & = (t_1-t_0) \frac{8\pi}{(d-2)}G M Q_S =0\,,
\end{split}
\label{ScalarChargeVanishingd}
\end{equation} 
which generalizes \eqref{ScalarChargeVanishing} to higher dimensions.

%%%%%%%%%%%%%%%%%%%%%%%%%%%%%%%%%%%%%%%%%%%%%%%%%%%%%%%%%
\section{Evading the no-hair theorem} 
\label{no_hair_violation}
In this section, we study 
how the 
no-hair theorem formulated in Sec.~\ref{SSST_Stationary} is affected if we drop some of its underlying assumptions. We will consider in particular two cases: a linear time dependence in the scalar profile, and a coupling to the Gauss--Bonnet operator, which violate the conditions (iv) and (v) respectively. We will also mention the possibility that assumption (i) is violated.

\subsection{Introducing a linear time dependence} 
\label{Time_Dependence}
 Let us start by relaxing the condition (iv). It is in general not strictly necessary for the scalar field solution $\psi$ to have the same symmetries as the background metric: one can in fact allow for a  linear dependence  in $t$ and/or  $\phi_a$ while keeping the isometries of the spacetime unchanged. 
 In practice, this happens because the shift symmetry ensures that the stress energy tensor depends on $\psi$ only through its derivatives. Said differently, although time translations and/or rotations in $\phi_a$ are spontaneously broken, it is possible to find, for each broken generator, a ``diagonal'' combination with a suitable  shift in $\psi$ that is unbroken on the background solution.
 
 It has been shown in \cite{Graham:2014ina} that a linear time dependence in the scalar profile is actually not allowed around stationary asymptotically flat black holes in the context of $K$-essence. However, the proof assumes a field $\psi$ that is backreacting on the metric through its energy-momentum tensor, while it is possible to find counterexamples to this statement in the test field limit. The simplest and most notable example of this is given by Jacobson's solution~\cite{Jacobson_1999} (see also~\cite{Charmousis:2019vnf}), where the scalar field carries a linear dependence on time.
More general examples of such ``stealth'' solutions beyond $K$-essence have been later found also in the context of e.g.~Schwarzschild-(A)dS black holes~\cite{Mukohyama:2005rw,Babichev:2013cya,Kobayashi:2014eva,BenAchour:2018dap,Motohashi:2019sen}
and Lorentz-violating gravity~\cite{Ramos:2018oku}. 

Given these preliminary considerations, let us consider a scalar profile of the form\footnote{For simplicity, we considered  only a linear term in $t$. Adding linear terms in the angles $\phi_a$ would not formally change our conclusions.}
\begin{equation}
    \psi = S(r,\theta)+E\,t \, ,
\label{psilineart}
\end{equation}
where we have included a linear dependence on time, and where we have denoted generically with $\theta$ the angles that do not correspond to Killing directions. Note that at large $r$, $S$ admits  the multipole expansion given by Eq.~\eqref{MultipolarExp_d}.

Let us start again from Eq.~\eqref{Stokes}.
The contributions to the integral from the fluxes through the hypersurfaces at $t=t_0$ and $t=t_1$ cancel out for the same reason as in the previous sections. A crucial difference is instead arising from the contribution at the horizon. To understand why, let us focus on the inequality in Eq.~\eqref{Cauchy_Schwarz}. In Sec.~\ref{higher_d} it was crucial that $J_t=J_{\phi_a}=0$ to be able to use the assumption (v) and conclude that  $J_i g^{ij} J_i$ is finite at the horizon. This is no longer true now, as a profile of the form in Eq.~\eqref{psilineart} will in general induce a nonzero $J_t$, allowing $J_i g^{ij} J_i$ to  be singular at $r=r_{\rm H}$ without invalidating the regularity of  $J^\mu J_\mu$. 

An example of this is given by Ref.~\cite{Jacobson_1999}.
That solution is valid in the limit in which the scalar field backreaction on the geometry is neglected; i.e.~it is an exact solution of the Klein--Gordon equation $\Box \psi = 0$ on a Schwarzschild background. It reads explicitly
\begin{equation}
    \psi(t,r) = Q_S\left[\frac{t}{r_{\rm H}}+\ln\Big(1-\frac{r_{\rm H}}{r}\Big)\right]\,,
\label{JacobsonScalar}
\end{equation}
where $r_{\rm H}$ denotes the Schwarzschild radius.  
The conserved shift-symmetry current in this case is simply $J^{\mu}=\partial^{\mu}\psi$. The squared norm $J^\mu J_\mu $ is thus just the standard kinetic term $\partial_\mu\psi\partial^\mu\psi$, which is regular at $r=r_{\rm H}$, as can  easily be verified using Eq.~\eqref{JacobsonScalar}. 
However, computing the left-hand side of Eq.~(\ref{Cauchy_Schwarz}) explicitly, we see that it does not vanish at $r=r_{\rm H}$. In fact,
\begin{equation}
     \int_{r=r_{\rm H}}{\rm d}\Sigma_{r}\,g^{rr}\,J_r = -4\pi \,r_H\,Q_S\,,
     \label{horizon_Jacobson}
\end{equation}
which is nonzero, as a result of the linear time dependence in Eq.~\eqref{JacobsonScalar}. Furthermore, the surface integral at $r_{\text{out}}\rightarrow \infty$ yields a contribution
equal in magnitude but with opposite sign. Therefore, Stokes' theorem is trivially satisfied and cannot be used to constrain the scalar charge $Q_S$.  
There are cases in which one can also have nontrivial contributions from the linear time dependence at $r = r_{\text{out}}$. 
To see this, let us study the asymptotic behavior of the current at large radii, keeping in mind the asymptotic expansion of the metric introduced in assumption (ii). The kinetic term now reads 
\begin{equation}
\begin{split}
    X &= -\frac{1}{2}\bigg[-E^2+(\partial_r S)^2+\frac{(\partial_{\theta} S)^2}{r^2}  \\ 
    &\qquad\qquad+h^{\alpha\beta}\partial_{\alpha} \psi\,\partial_{\beta} \psi\,\bigg] = \\
    &=-\frac{1}{2}\left[-E^2+h^{\alpha\beta}\partial_{\alpha}\psi\,\partial_{\beta}\psi+\mathcal{O}\left(\frac{1}{r^{2d-4}}\right)\right]\,,
\end{split}
\end{equation}
and its derivative is given by
\begin{equation}
\begin{split}
    \partial_rX &= -\frac{1}{2}\partial_r\left[-(1-h^{tt})E^2+\mathcal{O}\left(\frac{1}{r^{2d-4}}\right)\right] =\\
    &= -\frac{E^2}{2}\partial_rh^{tt}+\mathcal{O}\left(\frac{1}{r^{2d-3}}\right)\,.
\end{split}
\end{equation}
Regardless of its full expression, the metric component $h_{tt}$ at large distances must yield the Newtonian potential in $d$-dimensions, i.e.,
\begin{equation}
    \partial_rh^{tt}\sim \mathcal{O}\Big(\frac{1}{r^{d-2}}\Big)\,.
\end{equation}
It is thus clear that another important difference with the time-independent case [c.f.~Eq.~(\ref{asymptotic_current})] is showing up at large radii:
the gradient of the kinetic term now yields an additional contribution of the same order as the gradient of the scalar field. Note that this effect can  arise if the theory includes a cubic Galileon interaction. In the case of \eqref{Horndeski_current}, the current is  given asymptotically  by
\begin{equation}
\begin{split}
    J_r &= -\partial_rS\,(K_X+\mathcal{O}(1/r^{d-1}))\\
    &\quad -\partial_rX\,(G_{3X}+\mathcal{O}(1/r^{d-1}))\\
    &\quad\quad +\mathcal{O}(1/r^{d+1})\,.
    \label{J_r_SS}
\end{split}
\end{equation}
The Stokes' theorem then yields
\begin{equation}
\begin{split}
    &\oint_{\partial\mathcal{M}} {\rm d} \Sigma_{\mu}\,g^{\mu\nu}\,J_{\nu}= 0 =\\
    &=\int_{r=r_{\rm H}}{\rm d}\Sigma_{r}\,g^{rr}\,J_r+\int_{\{r\rightarrow\infty\}}{\rm d}\Sigma_{r}\,g^{rr}\,J_r = \\
    &=\int_{r=r_{\rm H}}{\rm d}\Sigma_{r}\,g^{rr}\,J_r\\
    &\quad-\int_{\{r\rightarrow\infty\}}{\rm d}\Sigma_{r}\Big(\partial_r S\,K_X+ \partial_rX\,G_{3X}\Big)\,.
\end{split}   
\label{ScalarChargeConstraint}
\end{equation}
Let us now specialize to $d = 4 $ dimensions and assume
\begin{equation}
    \begin{split}
        &K_X\rightarrow 1 \\
        &G_{3X}\rightarrow g_3\,
    \end{split}
\label{KXG3X}
\end{equation}
at large distances, as expected from asymptotic flatness~\cite{Barausse_Yagi}, where $g_3$ is the coupling associated with the cubic Galileon interaction. Then,  Eq.~(\ref{ScalarChargeConstraint}) provides one with a general expression for the scalar charge, 
\begin{equation}
    \begin{split}
        Q_S=g_3\,E^2-\frac{1}{4\pi\,GM\,M_{\rm Pl}^2}\int_{r=r_{\rm H}}{\rm d}\Sigma_{r}\,g^{rr}\,J_r\,.
    \end{split}
\label{scalar_charge_expression}
\end{equation}
Depending on the specific model, this constraint can be verified trivially (as in the case of Jacobson's solution), or it can yield nontrivial relations between the scalar charge and the scalar time gradient $E$. In particular, this second possibility  can give rise to hairy solutions 
in the presence of cubic Galileon interactions, if the  horizon contribution vanishes. 

\subsection{Coupling to Gauss-Bonnet}

It is well known that a linear coupling of the scalar field $\psi$ to the Gauss-Bonnet invariant $\mathcal{G} \equiv R^{\mu\nu\rho\sigma}R_{\mu\nu\rho\sigma} - 4R^{\mu\nu}R_{\mu\nu} + R^2$ can source a nontrivial hair around spherically symmetric black hole solutions in $d=4$, while preserving the shift symmetry \cite{General_Shift_Symmetric}.
In this case, the assumption that is violated is (v). In fact, the squared norm $J^\mu J_\mu$ of the shift-symmetry current  diverges at the horizon. This is however not an issue since $J^\mu$ is not a diffeomorphism-invariant current in the presence of the Gauss--Bonnet term, and therefore $J^\mu J_\mu$ is not a physical scalar quantity~\cite{Creminelli_2020}.

The same conclusion is expected to hold for rotating solutions. Note that the divergence of $J^\mu J_\mu$ at the horizon prevents one from claiming that the right-hand side of Eq.~(\ref{Cauchy_Schwarz_4d}) is zero at $r=r_{\rm H}$, invalidating our no-hair theorem of Sec.~\ref{SSST_Stationary}.

\subsection{Deviations from circularity}
We conclude this section with an additional cautionary note. The first assumption that we made was circularity, i.e.~we required that besides possessing two Killing vectors, the metric is also reflection symmetric. The existence of physically meaningful noncircular stationary, asymptotically flat rotating black holes seems to be excluded for a wide class of theories \cite{Nakashi_2020}. %\eb{changed, pls check } 
In an EFT context, Ref.~\cite{Xie_2021} showed that black hole solutions must be circular if they reduce to  GR solutions in a proper limit (in other words, if there are no separate branches). 

In principle, however, in theories beyond GR  one should not take circularity for granted~\cite{Anson_2021}. 
Evidence for deviations from circularity is found numerically by Ref.~\cite{Van_Aelst_2020} in cubic Galileon gravity. Other cases can be found for instance in  DHOST theories~\cite{Achour_2020, Takamori:2021atp}. These solutions involve a stealth time dependence of the scalar field, which does not show up in the Einstein equations. Furthermore, separate branches of solutions could exist in the presence of a coupling between the scalar and a curvature invariant like the Gauss-Bonnet term. (We will discuss this case in more detail in Sec. \ref{Quasi_Kessence}.)

In summary, while examples of hairy black holes that violate our assumption (i) do exist, they also seem to violate 
our assumptions (iv) or (v).

%%%%%%%%%%%%%%%%%%%%%%%%%%%%%%%%%%%%%%%%%%%%%%%%%%%%%%%
\section{No-hair theorem for quasi-$K$-essence theory}
\label{Quasi_Kessence} 
In this section, we review the no-hair theorem of Ref.~\cite{NoHair_KEssence},
which shows that in $K$-essence theories
the scalar field must be trivial
 and generalize it to generic spacetime dimensions $d$. For simplicity, we again restrict the analysis to the class of Myers-Perry black holes, with single or multiple spins. We stress that the proof is stronger than the one discussed so far, in the sense that it rules out not only a nonvanishing scalar charge at infinity, but
 also a nontrivial scalar profile. It is, however, less general, as it applies only to $K$-essence theories, i.e.~ones
 with first-order derivative self-interactions.
 In more general theories, the presence of higher derivative operators may invalidate the proof, as it will become clear later on. Nevertheless, 
 we will show that if
the higher derivative operators provide perturbative corrections to the $K$-essence action,
the proof can be generalized (with some subtle caveats).
 We will discuss this aspect at the end of the section. 

Let us start from the action\footnote{As opposed to the previous sections, we relax here the assumption of shift symmetry, allowing $K$ to be a function of both $X$ and $\psi$.}  
\begin{equation}
 S= \int {\rm d}^d x   \sqrt{-g} \, \frac{M_{\rm Pl}^2}{2} \left[R+ K(\psi,X)\right]\,,
\end{equation}
where $X\equiv -\frac{1}{2}\nabla_\mu\psi\nabla^\mu\psi$. 
Let us assume that there may exist a stationary black hole solution with nontrivial scalar profile, sharing the same symmetries as the geometry. The scalar equations of motion are
\begin{equation}
\nabla_{\alpha}\big(K_X\,\partial^{\alpha}\psi\big)+K_{\psi}= 0\,.
    \label{KEssence_EOM}
\end{equation}
Now, let us multiply Eq.~(\ref{KEssence_EOM}) by $\psi$ and integrate it over the volume of the black hole exterior region $\mathcal{M}$. Then, integrating by parts and  using  Stokes' theorem, we obtain
\begin{equation}
     \int_{\mathcal{M}}{\rm d}^dx\,\sqrt{-g}\left(K_X\,\partial_{\alpha}\psi\partial^{\alpha}\psi-\psi\,K_\psi\right) = \int_{\partial\mathcal{M}}{\rm d}\Sigma_{\alpha}\,V^{\alpha}
\label{Integrated_Kessence_EOM}
\end{equation}
where we have introduced the vector
\begin{equation}
    V^{\alpha}\equiv K_X\,\psi\,\partial^{\alpha}\psi\,.
\end{equation}
Let us focus first on the right-hand side of Eq.~\eqref{Integrated_Kessence_EOM}. 
The treatment of the surface terms is analogous to our previous discussion in Sec. \ref{SSST_Stationary}: the flux at the horizon is zero because of the vanishing of ${\rm d}\Sigma_{\alpha}{\rm d}\Sigma^{\alpha}$ and the regularity of $V_\alpha V^\alpha$ at $r=r_{\rm H}$; the contributions from the integral over the time boundaries cancel out because $V_\alpha$ is time independent; the flux across the hypersurface $r=r_{\text{out}}\rightarrow \infty $ is zero because we are considering asymptotically flat solutions and hence $\psi\rightarrow 0$ at large radii.  
As a result, the right-hand side of Eq.~(\ref{Integrated_Kessence_EOM}) vanishes. 

Let us then focus on the terms on the left-hand side of Eq.~\eqref{Integrated_Kessence_EOM}.
For a $\psi$ solution that has the same symmetries as the  geometry, we can write $\partial_{\alpha}\psi\partial^{\alpha}\psi=\partial_{i}\psi \, g^{ij}\partial_{j}\psi$, which is positive definite outside the horizon [see the discussion around Eqs.~\eqref{gdecomp}--\eqref{Sherman_Morrison}]. Furthermore, the energy-momentum tensor of the scalar field is $T_{\mu\nu}= K_X \partial_\mu\psi\partial_\nu\psi+K g_{\mu\nu}$. Assuming the null energy condition, i.e.~$n^\mu T_{\mu\nu}n^\nu\geq0$ for any null vector $n^\mu$
\cite{Rubakov:2014jja,Franciolini:2018aad}, one obtains $K_X\geq0$, which means that the term $K_X\,\partial_{\alpha}\psi\partial^{\alpha}\psi$ in Eq.~\eqref{Integrated_Kessence_EOM} is positive (semi)definite. Then, unless $\psi K_\psi$ in Eq.~\eqref{Integrated_Kessence_EOM} is also positive (semi)definite,\footnote{Note that, for $K=X-V(\psi)$, requiring $\psi K_\psi=-\psi V_\psi$ to be positive semidefinite is equivalent to having a potential $V$ that is unbounded from below.} the left-hand side of Eq.~\eqref{Integrated_Kessence_EOM} can vanish only if $\psi$ is the trivial solution \cite{NoHair_KEssence}.

Now, starting from this result, let us add a cubic Galileon term to the  Lagrangian:
\begin{equation}
    \frac{\Delta\mathcal{L}}{\sqrt{-g}}= -\frac{M_{\rm Pl}^3}{\Lambda^3}G_3(\psi,X)\square\psi\,,
\label{cubic_galileon}
\end{equation}
Note that in Eq.~(\ref{horndeski}) the energy scales $M_{\rm Pl}$ and $\Lambda$ are absorbed in the definition of $G_3$, while we show them explicitly here. With this definition, $G_3$ has the dimensions of an energy squared, like $X$ and $K$. 

The dynamics of the system is associated with a characteristic scale, set by the mass of the black hole, $M$. Let us therefore rescale the coordinates as $x^{\mu}\rightarrow x^{\mu}M_{\rm Pl}^2/M$. The scalar equation of motion then reads
\begin{equation}
\begin{split}
   \nabla_{\alpha}\big(K_X\,&\partial^{\alpha}\psi\big)+\Big(\frac{M}{M_{\rm Pl}^2}\Big)^2 \,K_{\psi}=\\
   &=\varepsilon \,\Big[\nabla_{\alpha}\big(G_{3X}\square\psi\,\partial^{\alpha}\psi+\,G_{3X}\,\partial^{\alpha}X \big)+\\
   &\qquad\qquad+\Big(\frac{M}{M_{\rm Pl}^2}\Big)^2\,G_{3\psi}\square\psi\Big]\,.
\end{split}   
\label{quasi_Kessence}
\end{equation}
where we defined the dimensionless parameter
\begin{equation}
    \varepsilon \equiv \frac{M_{\rm Pl}^5}{M^2\Lambda^3}\,.
\end{equation}
By requiring that the Galileon strong coupling scale $\Lambda$ is relevant for  the cosmological dynamics, one gets $\Lambda^3 \sim M_{\rm Pl}\,H^2$, with $H$  the Hubble parameter.\footnote{This can be found heuristically by equating the kinetic and Galileon energies, and estimating the derivatives with the Hubble rate $\partial\sim H$.} Then, the parameter $\varepsilon$ becomes the ratio of the Hubble  and  black hole  radii. This is of course a  huge number for astrophysical black holes. 
Therefore,  considering Galileon-like interactions in the scalar sector with a cutoff scale producing a non-negligible dynamics on cosmological scales, we can expect highly nonperturbative corrections to the $K$-essence solution $\psi =0$. 

However, one may consider the same interaction with a cutoff  $\Lambda$ large enough to make $\varepsilon \ll 1$ \cite{Noller:2019chl}. In this case, we will have small perturbative corrections to the $K$-essence scalar solution; i.e.~we can write
\begin{equation}
    \psi = \sum_{n=1}^{\infty}\varepsilon^n\psi^{(n)}\,.
    \label{perturbative_expansion}
\end{equation}
Plugging this ansatz into Eq.~(\ref{quasi_Kessence}),  at $\mathcal{O}(\varepsilon)$ one obtains
\begin{equation}
\label{perturbative_expansion2}
\nabla_{\alpha}\big(K_X(\psi=0)\,\partial^{\alpha}\psi^{(1)}\big)= 0\,,
\end{equation}
while the backreaction of the scalar field onto the metric through the Einstein equations is subleading. 
From this equation, one can conclude that $\psi^{(1)} = 0$. The procedure can be carried out iteratively for successive orders in $\varepsilon$. In more detail, at $\mathcal{O}(\varepsilon^n)$ one has
\begin{equation}
\begin{split}
    \nabla_{\alpha}\big(K_X(\psi=&0)\,\partial^{\alpha}\psi^{(n)}\big)+\\&+\mathcal{C}(\psi^{(1)},\dots,\psi^{(n-1)})= 0\,,
\end{split}
\end{equation}
where the corrections $\mathcal{C}$ vanish, as lower order corrections are zero.

Therefore, we can conclude that adding the  Galileon term  \eqref{cubic_galileon} in the action does not spoil the no-hair theorem proven for $K$-essence, as long as the coupling is perturbative. This holds straightforwardly also for the higher order Galileon interactions in the action (\ref{horndeski}).

The situation is different if we introduce, e.g., a coupling of the form
\begin{equation}
    \frac{\Delta\mathcal{L}}{\sqrt{-g}}=\alpha \,f(\psi)I
\end{equation}
where $I$ is a curvature invariant. The only invariants which can be relevant are the Ricci and the $m$th order Euler density, where $m=d/2$, $d$ being the dimension of the manifold. The former is zero on the trivial $K$-essence solution,
and therefore it can induce deviations from that solution only 
 at order $\alpha^2$. However, the latter is generally nonzero, even in vacuum, and can yield nontrivial contributions to the perturbative calculation. In $d=4$, the $m=2$ Euler density is the Gauss--Bonnet invariant
\begin{equation}
     \mathcal{G}= R^2-4R_{\alpha\beta}R^{\alpha\beta}+R_{\alpha\beta\mu\nu}R^{\alpha\beta\mu\nu}\,.
\end{equation}
In the following we will focus for simplicity on the case $I = \mathcal{G}$ in $d=4$, but the argument in a generic higher dimensional manifold would be completely analogous. 

The contribution to the scalar equation is 
\begin{equation}
    \frac{\delta}{\delta\psi}\Bigg(\frac{\Delta\mathcal{L}}{\sqrt{-g}}\Bigg)= \alpha_{\rm GB} f_{\psi}\,\mathcal{G}\,.
\end{equation}
One can rescale the coordinates 
as done earlier, and define the dimensionless parameter $\Tilde{\alpha}=\alpha_{\text{GB}}(M_{\rm Pl}^2/M)^2$. Carrying out the perturbative expansion with this parameter, one can see that the contribution to the equation of motion can be $\mathcal{O}(\Tilde{\alpha})$ if $f_{\psi}=\partial f/\partial \psi=\mathcal{O}(\Tilde{\alpha}^0)$. This is the case for shift-symmetric and dilatonic scalar-Gauss--Bonnet coupling functions, i.e.,
\begin{equation}
\begin{split}
    &f_{\text{Shift-Symm}}\sim\psi \, , \\
    &f_{\text{Dilatonic}}\sim {\rm e}^{\psi}\,.
\end{split}
\end{equation}
In these cases, one finds perturbative corrections to the $K$-essence solution $\psi=0$ at all orders. For instance, at first order, one has
\begin{equation}
    \nabla_{\alpha}\big(K_X(\psi=0)\,\partial^{\alpha}\psi^{(1)}\big)= -\mathcal{G}\,.
\end{equation}
This result is not surprising, as these coupling functions do not admit GR solutions, and the formation of a scalar hair is already evident at the perturbative level. 

There are, however, examples of coupling functions $f(\psi)$  admitting GR solutions, alongside ``scalarized'' ones \cite{No_Hair_GB,PhysRevLett.120.131103,Spin_Induced_SpontaneousScalarization,Spin_Induced_Scalarization_Herdeiro}.
Consider for instance black holes in scalar-Gauss-Bonnet gravity, with $f(\psi)=\psi^2$ and a canonical kinetic term $K(X)=X$. The scalar field obeys the Klein-Gordon equation
\begin{equation}
    \left(\Box+m_{\text{eff}}^2\right)\psi = 0\,,
\end{equation}
with the effective mass $m_{\text{eff}}^2=2\alpha_{\text{GB}}\, \mathcal{G}$.
The trivial GR configuration $\psi=0$ is clearly a solution, 
but nontrivial solutions  can also exist. In fact, $\mathcal{G}>0$
for the Schwarzschild solution and $\mathcal{G}<0$ for the Kerr one (at large spins). Therefore, depending on the sign of $\alpha_{\text{GB}}$, the effective mass can become tachyonic, in which case the GR solution 
is unstable. The instability's endpoint is a scalarized nontrivial solution. 

This solution is missed by our perturbative argument, which formally does apply to a theory with $f(\psi)=\psi^2$ and $K(X)=X$ (predicting incorrectly that scalarized solutions should not exist). The reason is that 
the scalarized solution  is 
a nonperturbative correction of the GR one,
i.e.~it lives on a different branch of solutions.
 For this reason, our perturbative proof, which assumes a small deviation from the $K$-essence solution, does not apply. 
 
 Another important caveat is that 
 black holes can dynamically evolve away from the GR configuration (even in the presence of  no-hair theorems).
 This is the case for instance for a nontachyonic mass term (produced by a quadratic coupling to the Gauss--Bonnet invariant, or simply by a scalar potential). For rotating black holes, a mass term can give rise to superradiant instabilities, i.e.~highly spinning GR solutions which, while allowed by no-hair theorems, may be unstable to superradiance~\cite{Damour:1976kh,Zouros:1979iw,Detweiler:1980uk,Dolan_2007,Dima_2020}.

%%%%%%%%%%%%%%%%%%%%%%%%%%%%%%%%%%%%%%%%%%%%%%%%%%%%%%%
\section{Conclusions}
 We proved a no-hair theorem for stationary, asymptotically flat black holes in shift-symmetric scalar-tensor theories. The theorem prevents the black holes from developing a scalar charge, defined as the coefficient of the $1/r$ falloff at large distances in the scalar profile. The proof is based on six fundamental assumptions and holds for general shift-symmetric theories, including Horndeski and DHOST theories as particular examples.\footnote{In the simplest case of a massless, noninteracting scalar field, note that the theorem is consistent with the symmetry arguments of \cite{Hui:2021vcv,Hui:2022vbh}.} Our result extends the no-hair theorem of Ref.~\cite{Hui:2012qt} to rotating black holes 
 %(although it is less general because it does not always exclude possible subleading falloff terms in the scalar profile)
 , as well as the results of Refs.~\cite{Bekenstein_I,Bekenstein_II,NoHair_KEssence}, which apply only to theories with at most single derivatives acting on the scalar field.
 
Under the assumption that the higher-derivative operators in the theory provide only perturbative corrections to the solution,  we showed, following \cite{NoHair_KEssence}, that there exists a stronger version of the theorem forbidding not only the scalar charge, but any nontrivial scalar profile.

Moreover, we discussed loopholes to the theorem, revisiting, in the context of rotating black holes, some known results in the literature. In particular, we discussed the case of a scalar field with linear dependence in time, and a coupling to the Gauss-Bonnet operator.

In addition, we showed how our no-hair theorem can be extended to higher spacetime dimensions in the class of Myers-Perry black holes with one or multiple spins. In this context, it would be interesting to study to what extent the theorem applies  also to rotating black holes with nontrivial topologies, which exist in $d>4$ \cite{Emparan:2008eg}. We leave this research direction for future work.

We conclude the discussion with a few remarks about the phenomenological implications of this result. 
Our theorem, unlike those of Refs.~\cite{Hui:2012qt,Bekenstein_I,Bekenstein_II,NoHair_KEssence}, does not  exclude possible subleading falloff terms in the scalar profile, unless we assume perturbative couplings. 
However, the absence of a scalar charge 
automatically proves that no deviations from
GR are to be expected in the gravitational wave fluxes at leading (i.e., dipole) order~\cite{ScalarCharge_Constraints}.

Furthermore, given the structure of Eq.~(\ref{scalar_charge_expression}), it would be interesting to find models of hairy black holes with linear time growth yielding a vanishing horizon term. In this case, the scalar charge would be independent of the geometry, and it would allow one to constrain the coupling $g_3$, assuming a growth rate for the scalar field at infinity comparable to the Hubble rate.

\section*{Acknowledgements}
We thank Christos Charmousis, Marco Crisostomi, Eric Gourgoulhon, Áron Kovács, Giulio Neri and Alessandro Podo for discussions and comments on a preliminary version of this manuscript.
LC and  EB acknowledge support from the European Union's H2020 ERC Consolidator Grant ``GRavity from Astrophysical to Microscopic Scales'' (Grant No.  GRAMS-815673). This work was also supported by the EU Horizon 2020 Research and Innovation Programme under the Marie Sklodowska-Curie Grant Agreement No. 101007855. LS is supported by the Centre National de la Recherche Scientifique (CNRS).

\appendix
\section{CONSERVED CURRENT AND CIRCULAR SPACETIMES}
\label{proof_Jt_0}
We claimed in Sec. \ref{SSST_Stationary} that the components $J_t$ and $J_\phi$ of the Noether current $J_\mu$, associated with the shift symmetry in the scalar action, are zero in our setup. Although not obvious, due to the presence in $J_\mu$ of possible higher derivative operators involving the curvature, it is not hard  to show that this is in fact the case as long as the conditions (i) and (iv) are fulfilled.

We can express the shift-symmetry current in full generality as
\begin{equation}
\label{current_form}
    J_{\mu}=F_{\mu\nu}\,\nabla^{\nu}\psi\,,
\end{equation}
where the tensor $F_{\mu\nu}$ is a function of $\psi$, $g_{\mu\nu}$ and their derivatives.
The idea of the proof, which we  formalize better in the following, consists simply in showing that, given $\nabla^t\psi =\nabla^\phi\psi=0$, it is not possible to construct (using only the building blocks $\psi$, $g$ and $\nabla$) any nontrivial $F_{\mu\nu}$ with an odd number of indices $r$ or $\theta$ if the spacetime has the isometries in (i).\footnote{A similar construction can be found, e.g., in Refs.~\cite{Franciolini:2018uyq,Hui:2021cpm}.}

Consider an open subset $\mathcal{U}$ of a four-dimensional spacetime, which has a continuous isometry group associated with two commuting Killing vectors, labeled as $\xi_{\mu}^{(a)}$, where $a = 1,2$. We define surface of transitivity \cite{doi:10.1063/1.1664763} a privileged two-dimensional hypersurface, which is everywhere tangent to the Killing vectors. A spacetime is circular if it has a set of Killing vectors and it admits hypersurfaces of conjugate dimension $2$, which are everywhere orthogonal to their transitivity surfaces (orthogonal transitivity condition). For a stationary circular black hole, we have the Killing vectors $\xi_{(a)}=\partial_t\,,\,\partial_\phi$. We can then define the surfaces of transitivity with the coordinates $i=r,\,\theta$ and choose the coordinates in such a way that the components of the metric $g_{ia}=g_{rt}\,,\,g_{r\phi}\,,\,g_{\theta t}\,,\,g_{\theta \phi}$ vanish. 

We then need a further step. Let us define $\zeta_{(i)}$ to be a set of independent vectors orthogonal to the transitivity surface. A tensor $T$ is said to be invertible at a point if all the quantities
\begin{equation}
    \tensor{T}{_{\mu_1,\,\dots\,\mu_p}^{\nu_1,\,\dots\,\nu_q}}\,\,\xi^{\mu_1}_{(a_1)}\,\dots\,\xi^{\mu_p}_{(a_p)}\,\zeta_{\nu_1}^{(i_1)}\,\dots\,\zeta_{\nu_q}^{(i_q)}
\end{equation}
are zero for odd $p$. 

It can be showed \cite{Xie_2021} that the circularity of the subset $\mathcal{U}$ implies the invertibility of the Riemann and Ricci tensors on it. Furthermore, it can also be proven that the invertibility property of a tensor is preserved when one takes its covariant derivative. 

Now consider Eq.~(\ref{current_form}). Given $\nabla^t\psi =\nabla^\phi\psi = 0$ and the symmetries of the Riemann tensor, it is easy to check that the only way of having a nonvanishing $J_t$ is to have nonzero tensors constructed from $R$ and covariant derivatives of $R$ and $\psi$, with an odd number of indices running over $(r,\theta)$. In other words, we should be able to construct a noninvertible $F_{\mu\nu}$ on $\mathcal{U}$. However, this cannot happen, thanks to the invertibility of the building blocks. 

The same can be shown for $J_\phi$. Hence, as long as the condition (iv) is verified, the only nonvanishing components of the current are $J_r$ and $J_\theta$.

Note that the situation is different if we drop the assumption (iv) and allow the scalar field to linearly depend on time. In this case, it is easy to verify that a non-vanishing $J_t$ is not incompatible with circularity anymore.

The same considerations can be extended to higher dimensions for the class of  black holes discussed in Sec.~\ref{higher_d}. 

\section{MYERS-PERRY BLACK HOLES IN $d$-DIMENSIONS}
\label{Kerr_d_dim}
The generalization of the Kerr metric in $d>4$ dimensions for a black hole rotating in a single plane is given by the Myers--Perry line element \cite{Myers_Perry} (see Ref.~\cite{Emparan:2008eg} for a review),
\begin{equation}
\begin{split}
    {\rm d}s^2 =& -{\rm d}t^2+\frac{\mu}{r^{d-5}\Sigma}({\rm d}t-a\,\sin^2\theta \,{\rm d}\phi)^2\\
    &+\frac{\Sigma}{\Delta}{\rm d}r^2+\Sigma \,{\rm d}\theta^2+(r^2+a^2)\sin^2\theta\,{\rm d}\phi^2\\
    &+r^2\cos^2\theta\,{\rm d}\Omega^2_{d-4}\,.
\end{split}
\label{Single_Plane_Metric}
\end{equation}
Besides the angles $\theta$ and $\phi$ defined in the usual way, we  have $d-4$ additional angles in ${\rm d}\Omega^2_{d-4}$. 
The functions appearing in the metric are generalizations of the well-known Kerr ones,
\begin{equation}
    \Sigma \equiv r^2+a^2\cos^2\theta\,, \quad\quad\quad \Delta \equiv r^2+a^2-\frac{\mu}{r^{d-5}}\,,
\end{equation}
with the mass and spin parameters being
\begin{equation}
    M \equiv \frac{(d-2)\Omega_{d-2}}{16\pi G}\mu\,, \quad\quad\quad J \equiv \frac{2M}{d-2}a\,\,.
\end{equation}
In $d>4$, rotation around more independent planes is allowed. In the most general case, the number of planes can be up to $N = (d-1)/2$. The metric is different for odd and even $d$. In particular, we have for odd $d$:
\begin{equation}
\begin{split}
    {\rm d}s^2 =& -{\rm d}t^2+(r^2+a_i^2)({\rm d}\mu_i^2+\mu_i^2{\rm d}\phi_i)+\\
     &+\frac{\mu r^2}{\Pi\,F}({\rm d}t-a_i\,\mu_i^2\,{\rm d}\phi_i)^2+\frac{\Pi\,F}{(\Pi-\mu r^2)}{\rm d}r^2\,,
\end{split}
\end{equation}
where $i=1,\dots,N$, $a_i$ is the spin associated to the $i$-th rotation plane and $\mu_i$ the corresponding direction cosine. Summation over $i$ is assumed and $\mu_i^2 = 1$. 

For even $d$ we have instead,
\begin{equation}
\begin{split}
     {\rm d}s^2 =& -{\rm d}t^2+r^2{\rm d}\alpha^2+(r^2+a_i^2)({\rm d}\mu_i^2+\mu_i^2{\rm d}\phi_i)+\\
     &+\frac{\mu r^2}{\Pi\,F}({\rm d}t-a_i\,\mu_i^2\,{\rm d}\phi_i)^2+\frac{\Pi\,F}{(\Pi-\mu r^2)}{\rm d}r^2\,.
\end{split}
\end{equation}
where the functions $\Pi$ and $F$ are defined as
\begin{equation}
\begin{split}
    &\Pi(r) = \prod_{I}^{N}(r^2+a_i^2) \, ,\\
    &F(r,\mu_i) = 1-\frac{a_i^2\mu_i^2}{r^2+a_i^2}\,,
\end{split}
\end{equation}
and $\mu_i^2+\alpha^2 = 1$.

\bibliographystyle{ieeetr}
\bibliography{Bibliography}

\begin{thebibliography}{100}

\bibitem{Fierz:1956zz}
M.~Fierz, ``{On the physical interpretation of P.Jordan's extended theory of
  gravitation},'' {\em Helv. Phys. Acta}, vol.~29, pp.~128--134, 1956.

\bibitem{Jordan:1959eg}
P.~Jordan, ``{The present state of Dirac's cosmological hypothesis},'' {\em Z.
  Phys.}, vol.~157, pp.~112--121, 1959.

\bibitem{Brans:1961sx}
C.~Brans and R.~H. Dicke, ``{Mach's principle and a relativistic theory of
  gravitation},'' {\em Phys. Rev.}, vol.~124, pp.~925--935, 1961.

\bibitem{Damour:1992we}
T.~Damour and G.~Esposito-Farese, ``{Tensor multiscalar theories of
  gravitation},'' {\em Class. Quant. Grav.}, vol.~9, pp.~2093--2176, 1992.

\bibitem{Damour:1993hw}
T.~Damour and G.~Esposito-Farese, ``{Nonperturbative strong field effects in
  tensor - scalar theories of gravitation},'' {\em Phys. Rev. Lett.}, vol.~70,
  pp.~2220--2223, 1993.

\bibitem{Will:1993hxu}
C.~M. Will, {\em {Theory and Experiment in Gravitational Physics}}.
\newblock Cambridge: Cambridge University Press, 1993.

\bibitem{Damour:1991rd}
T.~Damour and J.~H. Taylor, ``{Strong field tests of relativistic gravity and
  binary pulsars},'' {\em Phys. Rev.}, vol.~D45, pp.~1840--1868, 1992.

\bibitem{Will:2014kxa}
C.~M. Will, ``{The Confrontation between General Relativity and Experiment},''
  {\em Living Rev. Rel.}, vol.~17, p.~4, 2014.

\bibitem{Astier:2012ba}
P.~Astier and R.~Pain, ``{Observational Evidence of the Accelerated Expansion
  of the Universe},'' {\em Comptes Rendus Physique}, vol.~13, pp.~521--538,
  2012.

\bibitem{SupernovaCosmologyProject:1998vns}
S.~Perlmutter {\em et~al.}, ``{Measurements of $\Omega$ and $\Lambda$ from 42
  high redshift supernovae},'' {\em Astrophys. J.}, vol.~517, pp.~565--586,
  1999.

\bibitem{Planck:2015bue}
P.~A.~R. Ade {\em et~al.}, ``{Planck 2015 results. XIV. Dark energy and
  modified gravity},'' {\em Astron. Astrophys.}, vol.~594, p.~A14, 2016.

\bibitem{Planck:2018vyg}
N.~Aghanim {\em et~al.}, ``{Planck 2018 results. VI. Cosmological
  parameters},'' {\em Astron. Astrophys.}, vol.~641, p.~A6, 2020.
\newblock [Erratum: Astron.Astrophys. 652, C4 (2021)].

\bibitem{LIGOScientific:2016aoc}
B.~P. Abbott {\em et~al.}, ``{Observation of Gravitational Waves from a Binary
  Black Hole Merger},'' {\em Phys. Rev. Lett.}, vol.~116, no.~6, p.~061102,
  2016.

\bibitem{Nicolis:2008in}
A.~Nicolis, R.~Rattazzi, and E.~Trincherini, ``{The Galileon as a local
  modification of gravity},'' {\em Phys. Rev. D}, vol.~79, p.~064036, 2009.

\bibitem{Pirtskhalava:2015nla}
D.~Pirtskhalava, L.~Santoni, E.~Trincherini, and F.~Vernizzi, ``{Weakly Broken
  Galileon Symmetry},'' {\em JCAP}, vol.~09, p.~007, 2015.

\bibitem{Santoni:2018rrx}
L.~Santoni, E.~Trincherini, and L.~G. Trombetta, ``{Behind Horndeski:
  structurally robust higher derivative EFTs},'' {\em JHEP}, vol.~08, p.~118,
  2018.

\bibitem{Horndeski:1974wa}
G.~W. Horndeski, ``{Second-order scalar-tensor field equations in a
  four-dimensional space},'' {\em Int. J. Theor. Phys.}, vol.~10, pp.~363--384,
  1974.

\bibitem{Gleyzes:2014dya}
J.~Gleyzes, D.~Langlois, F.~Piazza, and F.~Vernizzi, ``{Healthy theories beyond
  Horndeski},'' {\em Phys. Rev. Lett.}, vol.~114, no.~21, p.~211101, 2015.

\bibitem{Zumalacarregui:2013pma}
M.~Zumalac\'arregui and J.~Garc\'\i{}a-Bellido, ``{Transforming gravity: from
  derivative couplings to matter to second-order scalar-tensor theories beyond
  the Horndeski Lagrangian},'' {\em Phys. Rev. D}, vol.~89, p.~064046, 2014.

\bibitem{Langlois:2015cwa}
D.~Langlois and K.~Noui, ``{Degenerate higher derivative theories beyond
  Horndeski: evading the Ostrogradski instability},'' {\em JCAP}, vol.~02,
  p.~034, 2016.

\bibitem{Langlois:2015skt}
D.~Langlois and K.~Noui, ``{Hamiltonian analysis of higher derivative
  scalar-tensor theories},'' {\em JCAP}, vol.~07, p.~016, 2016.

\bibitem{Crisostomi:2016tcp}
M.~Crisostomi, M.~Hull, K.~Koyama, and G.~Tasinato, ``{Horndeski: beyond, or
  not beyond?},'' {\em JCAP}, vol.~03, p.~038, 2016.

\bibitem{BenAchour:2016fzp}
J.~Ben~Achour, M.~Crisostomi, K.~Koyama, D.~Langlois, K.~Noui, and G.~Tasinato,
  ``{Degenerate higher order scalar-tensor theories beyond Horndeski up to
  cubic order},'' {\em JHEP}, vol.~12, p.~100, 2016.

\bibitem{BenAchour:2016cay}
J.~Ben~Achour, D.~Langlois, and K.~Noui, ``{Degenerate higher order
  scalar-tensor theories beyond Horndeski and disformal transformations},''
  {\em Phys. Rev. D}, vol.~93, no.~12, p.~124005, 2016.

\bibitem{Crisostomi:2016czh}
M.~Crisostomi, K.~Koyama, and G.~Tasinato, ``{Extended Scalar-Tensor Theories
  of Gravity},'' {\em JCAP}, vol.~04, p.~044, 2016.

\bibitem{Langlois:2018dxi}
D.~Langlois, ``{Dark energy and modified gravity in degenerate higher-order
  scalar\textendash{}tensor (DHOST) theories: A review},'' {\em Int. J. Mod.
  Phys. D}, vol.~28, no.~05, p.~1942006, 2019.

\bibitem{Noller:2019chl}
J.~Noller, L.~Santoni, E.~Trincherini, and L.~G. Trombetta, ``{Black Hole
  Ringdown as a Probe for Dark Energy},'' {\em Phys. Rev. D}, vol.~101,
  p.~084049, 2020.

\bibitem{Monitor:2017mdv}
B.~Abbott {\em et~al.}, ``{Gravitational Waves and Gamma-rays from a Binary
  Neutron Star Merger: GW170817 and GRB 170817A},'' {\em Astrophys. J. Lett.},
  vol.~848, no.~2, p.~L13, 2017.

\bibitem{Creminelli:2018xsv}
P.~Creminelli, M.~Lewandowski, G.~Tambalo, and F.~Vernizzi, ``{Gravitational
  Wave Decay into Dark Energy},'' {\em JCAP}, vol.~1812, p.~025, 2018.

\bibitem{Creminelli:2019nok}
P.~Creminelli, G.~Tambalo, F.~Vernizzi, and V.~Yingcharoenrat, ``{Resonant
  Decay of Gravitational Waves into Dark Energy},'' {\em JCAP}, vol.~10,
  p.~072, 2019.

\bibitem{Creminelli:2019kjy}
P.~Creminelli, G.~Tambalo, F.~Vernizzi, and V.~Yingcharoenrat, ``{Dark-Energy
  Instabilities induced by Gravitational Waves},'' {\em JCAP}, vol.~05, p.~002,
  2020.

\bibitem{Lara:2022gof}
G.~Lara, M.~Bezares, M.~Crisostomi, and E.~Barausse, ``{Robustness of kinetic
  screening against matter coupling},'' {\em Phys. Rev. D}, vol.~107, no.~4,
  p.~044019, 2023.

\bibitem{Barausse:2012da}
E.~Barausse, C.~Palenzuela, M.~Ponce, and L.~Lehner, ``{Neutron-star mergers in
  scalar-tensor theories of gravity},'' {\em Phys. Rev. D}, vol.~87, p.~081506,
  2013.

\bibitem{Palenzuela:2013hsa}
C.~Palenzuela, E.~Barausse, M.~Ponce, and L.~Lehner, ``{Dynamical scalarization
  of neutron stars in scalar-tensor gravity theories},'' {\em Phys. Rev. D},
  vol.~89, no.~4, p.~044024, 2014.

\bibitem{Armendariz-Picon:2000nqq}
C.~Armendariz-Picon, V.~F. Mukhanov, and P.~J. Steinhardt, ``{A Dynamical
  solution to the problem of a small cosmological constant and late time cosmic
  acceleration},'' {\em Phys. Rev. Lett.}, vol.~85, pp.~4438--4441, 2000.

\bibitem{Babichev:2009ee}
E.~Babichev, C.~Deffayet, and R.~Ziour, ``{k-Mouflage gravity},'' {\em Int. J.
  Mod. Phys. D}, vol.~18, pp.~2147--2154, 2009.

\bibitem{terHaar:2020xxb}
L.~ter Haar, M.~Bezares, M.~Crisostomi, E.~Barausse, and C.~Palenzuela,
  ``{Dynamics of Screening in Modified Gravity},'' {\em Phys. Rev. Lett.},
  vol.~126, p.~091102, 2021.

\bibitem{Bezares:2021yek}
M.~Bezares, L.~ter Haar, M.~Crisostomi, E.~Barausse, and C.~Palenzuela,
  ``{Kinetic screening in nonlinear stellar oscillations and gravitational
  collapse},'' {\em Phys. Rev. D}, vol.~104, no.~4, p.~044022, 2021.

\bibitem{Bezares:2021dma}
M.~Bezares, R.~Aguilera-Miret, L.~ter Haar, M.~Crisostomi, C.~Palenzuela, and
  E.~Barausse, ``{No Evidence of Kinetic Screening in Simulations of Merging
  Binary Neutron Stars beyond General Relativity},'' {\em Phys. Rev. Lett.},
  vol.~128, no.~9, p.~091103, 2022.

\bibitem{LIGOScientific:2021djp}
R.~Abbott {\em et~al.}, ``{GWTC-3: Compact Binary Coalescences Observed by LIGO
  and Virgo During the Second Part of the Third Observing Run},'' {\em arXiv
  e-prints}, 11 2021.

\bibitem{osti_4238685}
D.~M. Eardley, ``Observable effects of a scalar gravitational field in a binary
  pulsar,'' {\em Astrophys. J., Lett., v. 196, no. 2, pp. L59-L62}, 3 1975.

\bibitem{Will:1989sk}
C.~M. Will and H.~W. Zaglauer, ``{Gravitational Radiation, Close Binary
  Systems, and the Brans-dicke Theory of Gravity},'' {\em Astrophys. J.},
  vol.~346, p.~366, 1989.

\bibitem{Kerr:1963ud}
R.~P. Kerr, ``{Gravitational field of a spinning mass as an example of
  algebraically special metrics},'' {\em Phys. Rev. Lett.}, vol.~11,
  pp.~237--238, 1963.

\bibitem{Newman:1965my}
E.~T. Newman and A.~I. Janis, ``Note on the kerr spinning‐particle metric,''
  {\em Journal of Mathematical Physics}, vol.~6, no.~6, pp.~915--917, 1965.

\bibitem{Carter:1968rr}
B.~Carter, ``Global structure of the kerr family of gravitational fields,''
  {\em Phys. Rev.}, vol.~174, pp.~1559--1571, Oct 1968.

\bibitem{PhysRevLett.34.905}
D.~C. Robinson, ``Uniqueness of the kerr black hole,'' {\em Phys. Rev. Lett.},
  vol.~34, pp.~905--906, Apr 1975.

\bibitem{cmp/1103857885}
S.~W. Hawking, ``{Black holes in the Brans-Dicke theory of gravitation},'' {\em
  Communications in Mathematical Physics}, vol.~25, no.~2, pp.~167 -- 171,
  1972.

\bibitem{Star_NoHair_Horndeski}
A.~Leh{\'{e} }bel, E.~Babichev, and C.~Charmousis, ``A no-hair theorem for
  stars in horndeski theories,'' {\em Journal of Cosmology and Astroparticle
  Physics}, vol.~2017, pp.~037--037, jul 2017.

\bibitem{Challenging_Scalar_Charge}
K.~Yagi, L.~C. Stein, and N.~Yunes, ``Challenging the presence of scalar charge
  and dipolar radiation in binary pulsars,'' {\em Physical Review D}, vol.~93,
  jan 2016.

\bibitem{NoHair_KEssence}
A.~A. Graham and R.~Jha, ``Nonexistence of black holes with noncanonical scalar
  fields,'' {\em Physical Review D}, vol.~89, apr 2014.

\bibitem{Jacobson_1999}
T.~Jacobson, ``Primordial black hole evolution in tensor-scalar cosmology,''
  {\em Physical Review Letters}, vol.~83, pp.~2699--2702, oct 1999.

\bibitem{Charmousis:2019vnf}
C.~Charmousis, M.~Crisostomi, R.~Gregory, and N.~Stergioulas, ``{Rotating Black
  Holes in Higher Order Gravity},'' {\em Phys. Rev. D}, vol.~100, no.~8,
  p.~084020, 2019.

\bibitem{Stealth_Hairs}
K.~Takahashi and H.~Motohashi, ``General relativity solutions with stealth
  scalar hair in quadratic higher-order scalar-tensor theories,'' {\em Journal
  of Cosmology and Astroparticle Physics}, vol.~2020, pp.~034--034, jun 2020.

\bibitem{Van_Aelst_2020}
K.~V. Aelst, E.~Gourgoulhon, P.~Grandcl{\'{e}}ment, and C.~Charmousis, ``Hairy
  rotating black holes in cubic galileon theory,'' {\em Classical and Quantum
  Gravity}, vol.~37, p.~035007, jan 2020.

\bibitem{Hui:2012qt}
L.~Hui and A.~Nicolis, ``{No-Hair Theorem for the Galileon},'' {\em Phys. Rev.
  Lett.}, vol.~110, p.~241104, 2013.

\bibitem{sGB_4D}
D.~Glavan and C.~Lin, ``Einstein-gauss-bonnet gravity in four-dimensional
  spacetime,'' {\em Physical Review Letters}, vol.~124, feb 2020.

\bibitem{GaussBonnetGravity_Review}
P.~G.~S. Fernandes, P.~Carrilho, T.~Clifton, and D.~J. Mulryne, ``The 4d
  einstein{\textendash}gauss{\textendash}bonnet theory of gravity: a review,''
  {\em Classical and Quantum Gravity}, vol.~39, p.~063001, feb 2022.

\bibitem{General_Shift_Symmetric}
T.~P. Sotiriou and S.-Y. Zhou, ``{Black hole hair in generalized scalar-tensor
  gravity},'' {\em Phys. Rev. Lett.}, vol.~112, p.~251102, 2014.

\bibitem{General_Shift_Symmetric_2}
T.~P. Sotiriou and S.-Y. Zhou, ``Black hole hair in generalized scalar-tensor
  gravity: An explicit example,'' {\em Physical Review D}, vol.~90, dec 2014.

\bibitem{Dilaton_BH}
P.~Kanti, N.~E. Mavromatos, J.~Rizos, K.~Tamvakis, and E.~Winstanley,
  ``Dilatonic black holes in higher curvature string gravity,'' {\em Physical
  Review D}, vol.~54, pp.~5049--5058, oct 1996.

\bibitem{sGB_Gravity_Solutions}
M.~{G{\"u}rses}, ``{Some solutions of the Gauss Bonnet gravity with scalar
  field in four dimensions},'' {\em General Relativity and Gravitation},
  vol.~40, pp.~1825--1830, Sept. 2008.

\bibitem{Spin_Induced_Scalarization_Herdeiro}
C.~A. Herdeiro, E.~Radu, H.~O. Silva, T.~P. Sotiriou, and N.~Yunes,
  ``Spin-induced scalarized black holes,'' {\em Physical Review Letters},
  vol.~126, jan 2021.

\bibitem{Spin_Induced_SpontaneousScalarization}
A.~Dima, E.~Barausse, N.~Franchini, and T.~P. Sotiriou, ``Spin-induced black
  hole spontaneous scalarization,'' {\em Physical Review Letters}, vol.~125,
  dec 2020.

\bibitem{Goldberger:2004jt}
W.~D. Goldberger and I.~Z. Rothstein, ``{An Effective field theory of gravity
  for extended objects},'' {\em Phys. Rev. D}, vol.~73, p.~104029, 2006.

\bibitem{Porto:2016pyg}
R.~A. Porto, ``{The effective field theorist\textquoteright{}s approach to
  gravitational dynamics},'' {\em Phys. Rept.}, vol.~633, pp.~1--104, 2016.

\bibitem{Hui:2020xxx}
L.~Hui, A.~Joyce, R.~Penco, L.~Santoni, and A.~R. Solomon, ``{Static response
  and Love numbers of Schwarzschild black holes},'' {\em JCAP}, vol.~04,
  p.~052, 2021.

\bibitem{1973blho.conf..241B}
J.~M. {Bardeen}, ``{Rapidly rotating stars, disks, and black holes.},'' in {\em
  Black Holes (Les Astres Occlus)}, pp.~241--289, Jan. 1973.

\bibitem{darboux1910lecons}
G.~Darboux, {\em Le{\c{c}}ons sur les syst{\'e}mes orthogonaux et les
  coordonn{\'e}es curvilignes}.
\newblock Cours de geometrie de la facult{\'e} des sciences, GAuthier Villars,
  1910.

\bibitem{chandrasekhar1992mathematical}
S.~Chandrasekhar, {\em The Mathematical Theory of Black Holes}.
\newblock International series of monographs on physics, Oxford University
  Press, 1992.

\bibitem{PhysRevLett.26.331}
B.~Carter, ``Axisymmetric black hole has only two degrees of freedom,'' {\em
  Phys. Rev. Lett.}, vol.~26, pp.~331--333, Feb 1971.

\bibitem{Bekenstein_II}
J.~D. Bekenstein, ``Nonexistence of baryon number for black holes. ii.,'' {\em
  Phys. Rev., D 5: No. 10, 2403-12(15 May 1972).}, 1 1972.

\bibitem{Saravani_2019}
M.~Saravani and T.~P. Sotiriou, ``Classification of shift-symmetric horndeski
  theories and hairy black holes,'' {\em Physical Review D}, vol.~99, jun 2019.

\bibitem{Benkel_2018}
R.~Benkel, N.~Franchini, M.~Saravani, and T.~P. Sotiriou, ``Causal structure of
  black holes in shift-symmetric horndeski theories,'' {\em Physical Review D},
  vol.~98, sep 2018.

\bibitem{Constraints_GB_Gravity}
K.~Yagi, ``New constraint on scalar gauss-bonnet gravity and a possible
  explanation for the excess of the orbital decay rate in a low-mass x-ray
  binary,'' {\em Physical Review D}, vol.~86, oct 2012.

\bibitem{Bekenstein_I}
J.~D. Bekenstein, ``Nonexistence of baryon number for static black holes,''
  {\em Phys. Rev. D}, vol.~5, pp.~1239--1246, Mar 1972.

\bibitem{Emparan:2008eg}
R.~Emparan and H.~S. Reall, ``{Black Holes in Higher Dimensions},'' {\em Living
  Rev. Rel.}, vol.~11, p.~6, 2008.

\bibitem{Myers_Perry}
R.~Myers and M.~Perry, ``Black holes in higher dimensional space-times,'' {\em
  Annals of Physics}, vol.~172, no.~2, pp.~304--347, 1986.

\bibitem{Graham:2014ina}
A.~A.~H. Graham and R.~Jha, ``{Stationary Black Holes with Time-Dependent
  Scalar Fields},'' {\em Phys. Rev. D}, vol.~90, no.~4, p.~041501, 2014.

\bibitem{Mukohyama:2005rw}
S.~Mukohyama, ``{Black holes in the ghost condensate},'' {\em Phys. Rev. D},
  vol.~71, p.~104019, 2005.

\bibitem{Babichev:2013cya}
E.~Babichev and C.~Charmousis, ``{Dressing a black hole with a time-dependent
  Galileon},'' {\em JHEP}, vol.~08, p.~106, 2014.

\bibitem{Kobayashi:2014eva}
T.~Kobayashi and N.~Tanahashi, ``{Exact black hole solutions in shift symmetric
  scalar\textendash{}tensor theories},'' {\em PTEP}, vol.~2014, p.~073E02,
  2014.

\bibitem{BenAchour:2018dap}
J.~Ben~Achour and H.~Liu, ``{Hairy Schwarzschild-(A)dS black hole solutions in
  degenerate higher order scalar-tensor theories beyond shift symmetry},'' {\em
  Phys. Rev. D}, vol.~99, no.~6, p.~064042, 2019.

\bibitem{Motohashi:2019sen}
H.~Motohashi and M.~Minamitsuji, ``{Exact black hole solutions in
  shift-symmetric quadratic degenerate higher-order scalar-tensor theories},''
  {\em Phys. Rev. D}, vol.~99, no.~6, p.~064040, 2019.

\bibitem{Ramos:2018oku}
O.~Ramos and E.~Barausse, ``{Constraints on Ho\v{r}ava gravity from binary
  black hole observations},'' {\em Phys. Rev. D}, vol.~99, no.~2, p.~024034,
  2019.
\newblock [Erratum: Phys.Rev.D 104, 069904 (2021)].

\bibitem{Barausse_Yagi}
E.~Barausse and K.~Yagi, ``Gravitation-wave emission in shift-symmetric
  horndeski theories,'' {\em Physical Review Letters}, vol.~115, nov 2015.

\bibitem{Creminelli_2020}
P.~Creminelli, N.~Loayza, F.~Serra, E.~Trincherini, and L.~G. Trombetta,
  ``Hairy black-holes in shift-symmetric theories,'' {\em Journal of High
  Energy Physics}, vol.~2020, aug 2020.

\bibitem{Nakashi_2020}
K.~Nakashi and M.~Kimura, ``Towards rotating noncircular black holes in
  string-inspired gravity,'' {\em Physical Review D}, vol.~102, oct 2020.

\bibitem{Xie_2021}
Y.~Xie, J.~Zhang, H.~O. Silva, C.~de~Rham, H.~Witek, and N.~Yunes, ``Square peg
  in a circular hole: Choosing the right ansatz for isolated black holes in
  generic gravitational theories,'' {\em Physical Review Letters}, vol.~126,
  jun 2021.

\bibitem{Anson_2021}
T.~Anson, E.~Babichev, C.~Charmousis, and M.~Hassaine, ``Disforming the kerr
  metric,'' {\em Journal of High Energy Physics}, vol.~2021, jan 2021.

\bibitem{Achour_2020}
J.~B. Achour, H.~Liu, H.~Motohashi, S.~Mukohyama, and K.~Noui, ``On rotating
  black holes in {DHOST} theories,'' {\em Journal of Cosmology and
  Astroparticle Physics}, vol.~2020, pp.~001--001, nov 2020.

\bibitem{Takamori:2021atp}
Y.~Takamori, A.~Naruko, Y.~Sakurai, K.~Takahashi, D.~Yamauchi, and C.-M. Yoo,
  ``{Testing the Non-circularity of the Spacetime around Sagittarius A* with
  Orbiting Pulsars},'' {\em {}}, 8 2021.

\bibitem{Rubakov:2014jja}
V.~A. Rubakov, ``{The Null Energy Condition and its violation},'' {\em Phys.
  Usp.}, vol.~57, pp.~128--142, 2014.

\bibitem{Franciolini:2018aad}
G.~Franciolini, L.~Hui, R.~Penco, L.~Santoni, and E.~Trincherini, ``{Stable
  wormholes in scalar-tensor theories},'' {\em JHEP}, vol.~01, p.~221, 2019.

\bibitem{No_Hair_GB}
H.~O. Silva, J.~Sakstein, L.~Gualtieri, T.~P. Sotiriou, and E.~Berti,
  ``Spontaneous scalarization of black holes and compact stars from a
  gauss-bonnet coupling,'' {\em Physical Review Letters}, vol.~120, mar 2018.

\bibitem{PhysRevLett.120.131103}
D.~D. Doneva and S.~S. Yazadjiev, ``New gauss-bonnet black holes with
  curvature-induced scalarization in extended scalar-tensor theories,'' {\em
  Phys. Rev. Lett.}, vol.~120, p.~131103, Mar 2018.

\bibitem{Damour:1976kh}
T.~Damour, N.~Deruelle, and R.~Ruffini, ``{On Quantum Resonances in Stationary
  Geometries},'' {\em Lett. Nuovo Cim.}, vol.~15, pp.~257--262, 1976.

\bibitem{Zouros:1979iw}
T.~J.~M. Zouros and D.~M. Eardley, ``{INSTABILITIES OF MASSIVE SCALAR
  PERTURBATIONS OF A ROTATING BLACK HOLE},'' {\em Annals Phys.}, vol.~118,
  pp.~139--155, 1979.

\bibitem{Detweiler:1980uk}
S.~L. Detweiler, ``{KLEIN-GORDON EQUATION AND ROTATING BLACK HOLES},'' {\em
  Phys. Rev. D}, vol.~22, pp.~2323--2326, 1980.

\bibitem{Dolan_2007}
S.~R. Dolan, ``Instability of the massive klein-gordon field on the kerr
  spacetime,'' {\em Physical Review D}, vol.~76, oct 2007.

\bibitem{Dima_2020}
A.~Dima and E.~Barausse, ``Numerical investigation of plasma-driven
  superradiant instabilities,'' {\em Classical and Quantum Gravity}, vol.~37,
  p.~175006, aug 2020.

\bibitem{Hui:2021vcv}
L.~Hui, A.~Joyce, R.~Penco, L.~Santoni, and A.~R. Solomon, ``{Ladder symmetries
  of black holes. Implications for love numbers and no-hair theorems},'' {\em
  JCAP}, vol.~01, no.~01, p.~032, 2022.

\bibitem{Hui:2022vbh}
L.~Hui, A.~Joyce, R.~Penco, L.~Santoni, and A.~R. Solomon, ``{Near-zone
  symmetries of Kerr black holes},'' {\em JHEP}, vol.~09, p.~049, 2022.

\bibitem{ScalarCharge_Constraints}
E.~Barausse, N.~Yunes, and K.~Chamberlain, ``Theory-agnostic constraints on
  black-hole dipole radiation with multiband gravitational-wave astrophysics,''
  {\em Physical Review Letters}, vol.~116, jun 2016.

\bibitem{Franciolini:2018uyq}
G.~Franciolini, L.~Hui, R.~Penco, L.~Santoni, and E.~Trincherini, ``{Effective
  Field Theory of Black Hole Quasinormal Modes in Scalar-Tensor Theories},''
  {\em JHEP}, vol.~02, p.~127, 2019.

\bibitem{Hui:2021cpm}
L.~Hui, A.~Podo, L.~Santoni, and E.~Trincherini, ``{Effective Field Theory for
  the perturbations of a slowly rotating black hole},'' {\em JHEP}, vol.~12,
  p.~183, 2021.

\bibitem{doi:10.1063/1.1664763}
B.~Carter, ``Killing horizons and orthogonally transitive groups in
  space‐time,'' {\em Journal of Mathematical Physics}, vol.~10, no.~1,
  pp.~70--81, 1969.

\end{thebibliography}

\end{document}